\documentclass[%
 reprint,
 superscriptaddress,
 showkeys,
 amsmath,amssymb,
 aps, 
 prl,
floatfix,
]{revtex4-2}

\usepackage{graphicx}
\usepackage{dcolumn}
\usepackage{bm}
\usepackage{color}
\usepackage[T1]{fontenc}
\usepackage{mathptmx}
\usepackage{hyperref}
\usepackage{tabularx}
\begin{document}
\newcolumntype{L}[1]{>{\raggedright\arraybackslash}p{#1}}
\newcolumntype{C}[1]{>{\centering\arraybackslash}p{#1}}
\newcolumntype{R}[1]{>{\raggedleft\arraybackslash}p{#1}}
\title{Linking continuum-scale state of wetting to pore-scale contact angles in porous media}
\author{Chenhao Sun}
\affiliation{School of Minerals \& Energy Resources Engineering, University of New South Wales, Kensington, NSW 2052, Australia}
\author{James E. McClure}
\affiliation{Advanced Research Computing, Virginia Polytechnic Institute \& State University, Blacksburg, Virginia 24061, USA}
\author{Peyman Mostaghimi}
\affiliation{School of Minerals \& Energy Resources Engineering, University of New South Wales, Kensington, NSW 2052, Australia}
\author{Anna L. Herring}
\affiliation{Department of Applied Mathematics, Australian National University, Canberra, ACT 2600, Australia}
\author{Mehdi Shabaninejad}
\affiliation{Department of Applied Mathematics, Australian National University, Canberra, ACT 2600, Australia}
\author{Steffen Berg}
\affiliation{Rock \& Fluid Physics, Shell Global Solutions International B.V., Grasweg 31, 1031 HW Amsterdam, The Netherlands}
\affiliation{Department of Earth Science \& Engineering, Imperial College London, London SW7 2AZ, UK}
\affiliation{Department of Chemical Engineering, Imperial College London, London SW7 2AZ, UK}
\author{Ryan T. Armstrong}
\email{ryan.armstrong@unsw.edu.au}
\affiliation{School of Minerals \& Energy Resources Engineering, University of New South Wales, Kensington, NSW 2052, Australia}

\begin{abstract}
Wetting phenomena play a key role in flows through porous media. Relative permeability and capillary pressure-saturation functions show a high sensitivity to wettability, which has different definitions at the continuum- and pore-scale. At the continuum-scale, the state of wetting is defined as Amott-Harvey or USBM (United States Bureau of Mines) indices by capillary pressure drainage and imbibition cycles. At the pore-scale, the concept of contact angle is used, which until recently was not experimentally possible to determine within an opaque porous medium. Recent progress on measurements of pore-scale contact angles by X-ray computed micro-tomography has therefore attracted significant attention in various research communities. In this work, the Gauss-Bonnet theorem is applied to provide a direct link between capillary pressure saturation  ($P_c(S_w)$) data and measured distributions of pore-scale contact angles. We propose that the wetting state of a porous medium can be described in terms of geometrical arguments that constrain the morphological state of immiscible fluids. The constraint describes the range of possible contact angles and interfacial curvatures that can exist for a given system. We present measurements in a tested sandstone for which the USBM index, $P_c(S_w)$, and pore-scale contact angles are measured. Additional studies are also performed using two-phase Lattice Boltzmann simulations to test a wider range of wetting conditions. We show that mean pore-scale contact angle measurements can be predicted from petrophysical data within a few differences. This provides a general framework on how continuum-scale data can be used to describe the geometrical state of fluids within porous media. 
\end{abstract}

\maketitle

\section{1. Introduction}

In a multiphase system, wettability refers to the relative preference between two fluids to coat solid materials as a consequence of associated interfacial energies \cite{bonn2009wetting,de1985wetting}. This physical consequence is important for numerous engineered systems and technological applications, such as microfluidics, hydrologic systems, hydrocarbon recovery, aquifer storage, printing and self-cleaning technologies \cite{Weitz2007Dripping,bauters2000physics,morrow1970physics,botto2017effects,PaulGatenholm1990Wetting,blossey2003self}. In porous media, it influences the physical processes of multiphase flow, which involves relative permeability, capillary pressure, capillary-end effect and counter-current imbibition \cite{anderson1987wettability,anderson1986wettability,kovscek1993pore,donaldson1969wettability}. While various laboratory methods can be employed to define wettability, there lacks a fundamental framework to incorporate it into continuum-scale models. Over the past few decades, phenomenological metrics, such as the Amott-Harvey or USBM (United States Bureau of Mines) indices are applied to define the wetting state of porous media \cite{donaldson1969wettability,anderson1987wettability}. More recent imaging advances have allowed for direct pore-scale contact angle measurements to represent \textit{in situ} wetting condition of the system \cite{Prodanovic2006,andrew2014pore,alratrout2017automatic,scanziani2017automatic}. These measurements bring about an exciting new tool to study wetting behavior in porous media. However, the tested systems provide a broad distribution of contact angles, which raises various fundamental questions regarding the physical reasons and uniqueness of these measurements.

Two commonly used wettability characterization methods should be considered to discern what are missing from our current understandings: (1) a macroscopic perspective based on the internal energy of the porous medium and (2) a microscopic perspective linking surface energies and contact angle. These methods provide similar qualitative information while in practice they are fundamentally different by measuring wettability at different length scales. The industry practice to quantify the wettability of a porous material is performed by measuring its capillary pressure versus saturation ($P_c(S_w)$) relationship \cite{mcphee2015wettability}. From this measurement, a wettability index can be defined, such as the Amott-Harvey or USBM indices \cite{amott1959observations,donaldson1969wettability}. For instance, the USBM index is defined as the ratio of the areas under the $P_c(S_w)$ curve between: (1) the saturation range that provides only positive capillary pressures and (2) the saturation range that provides only negative capillary pressures. Assuming that saturation change is an isothermal, reversible process with minimal dissipative processes, the area under the $P_c(S_w)$ curve for a given range of saturation can then be related to the work associated with that change \cite{donaldson1969wettability,morrow1970physics}. Therefore, the USBM index provides a ratio between the positive and negative work required for a porous medium to become saturated with a given fluid. This measurement provides a single bulk appraisal of average behaviour, which can be considered as a continuum-scale metric for wettability. However, there is no 
transparent mechanism to link such measures with fundamental wetting phenomena that can be observed at smaller scales.

Pore-scale measurements provide the contact angle formed between the fluids and solid defined by Young's equation \cite{young1805iii}, 
\begin{equation}
\sigma_{as} = \sigma_{bs} + \sigma_{ab} \cos(\theta),
\label{Modified_YL}
\end{equation}
where $\theta$ is the angle formed along the contact line in the orthogonal plane when the tangential forces caused by the interfacial ($\sigma_{ab}$, $\sigma_{as}$) and surface tensions ($\sigma_{bs}$) are in balance at equilibrium. This measurement is not necessarily practical for the continuum-scale quantification of wettability. The measurement of $\theta$ in porous media requires experiments at \textit{in situ} conditions and X-ray micro-computed tomography (micro-CT) technique followed by a sequence of image processing steps that introduce user-biases and have their own inherent difficulties with precision and accuracy. The measurement that depends on image resolution is taken only along the three-phase contact regions along the solid surface. The variations of surface roughness and chemical heterogeneity on the solid surface that lead to a wide variation of interfacial energies are not necessarily quantified adequately. In addition, the location of the contact line and observed $\theta$ are dependent on the system parameters. A high versus low capillary number experiment would push the contact line into different regions of the pore space. Drainage versus imbibition experiments would produce various distributions of advancing and receding contact angles \cite{morrow1975effects}. Contact angle hysteresis, interface pinning, and the time allotted for the system to reach equilibrium prior to imaging would also influence the observed pore-scale contact angles \cite{cassie1944wettability,morrow1975effects,indekeu1994line,heslot1990experiments,schaffer2000contact}.

To date, there exists no fundamental well-defined link between the two aforementioned measurements. In this work, we aim to provide a transparent link between pore-scale contact angle measurements and continuum-scale wettability indices. This is accomplished by considering the geometric state of fluid clusters in a porous medium, which is described by the relationship of local interfacial curvatures to global topology, as stated in the acclaimed Gauss-Bonnet theorem. It allows us to describe all possible geometrical arrangements of an fluid cluster in a porous medium with an assigned probability. The states and assigned probabilities are then related to the distribution of contact angle measurements commonly reported for multiphase systems in porous media. The underlying assumption is that capillary pressure, $P_c$, is a consequence of the surface roughness, chemical heterogeneity and pore sizes and that $P_c(S_w)$ data provides an adequate description of all possible capillary pressures. This assumption will be tested for rough sandstone domains with various distributions of surface energies.

\section{2. Theoretical Development}

\begin{figure}
\includegraphics[width=0.47\textwidth]{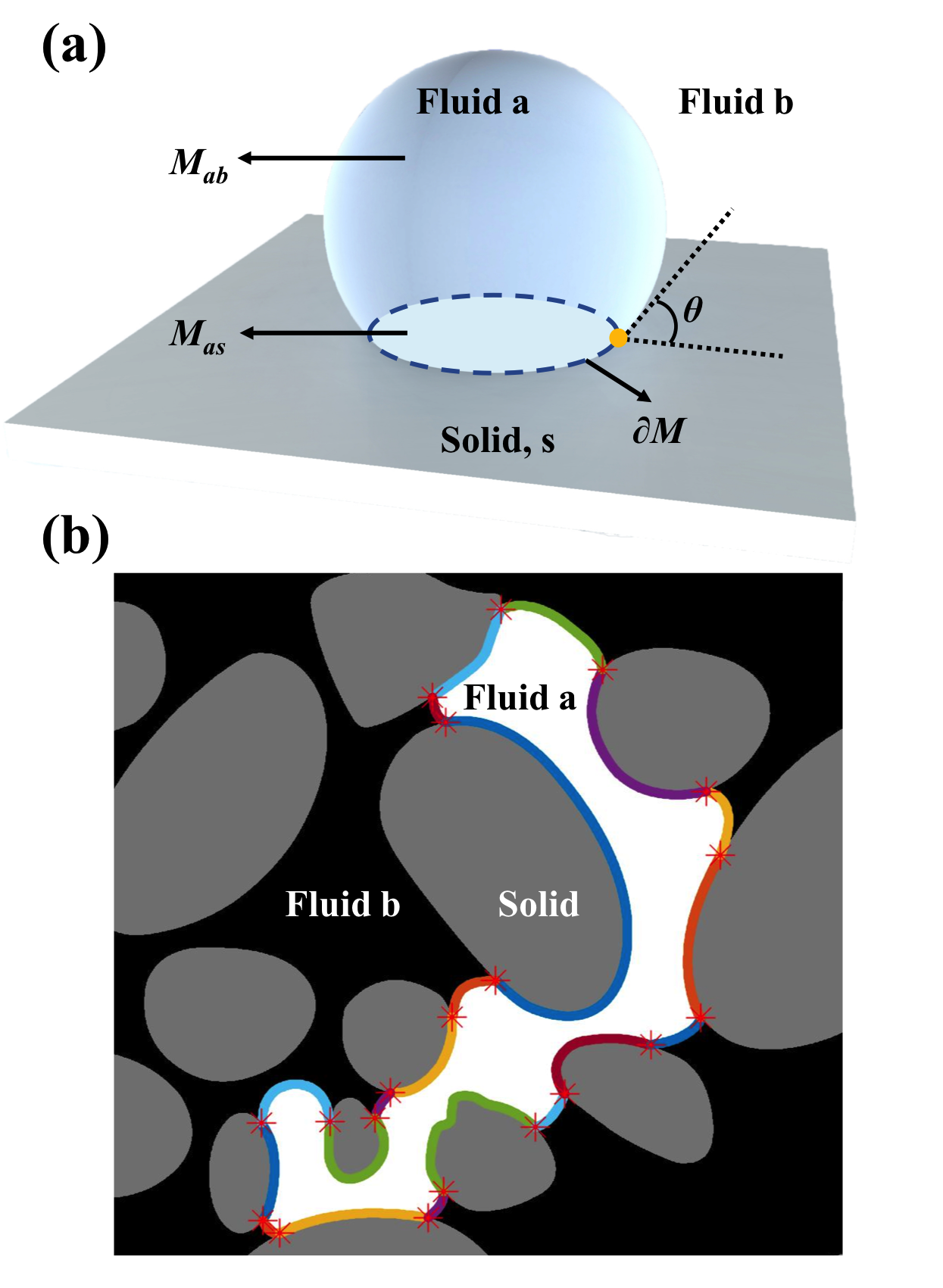}
\caption{(a) A three-dimensional (3D) schematic of a sessile droplet deposited on solid surface immersed within an immiscible fluid. The contact line, $\partial M$, separates the boundary surface of the cluster into two interfaces, $M_{ab}$ and $M_{as}$ respectively. (b) A two-dimensional (2D) segmented image that shows a cluster in real porous medium where the surface is deformed by grains.}
\label{fig1}
\end{figure}

The Gauss-Bonnet theorem can be applied to relate the total Gaussian curvature of a fluid cluster ($C$) to its global topology \cite{chern1944simple}, 

\begin{align}
2 \pi \chi(C) = \int_{M} \kappa_T dM+ \int_{\partial M} \kappa_g dC,
\label{GBT}	
\end{align}
where $dM$ is an area element along the cluster surface. $\kappa_T = 1/(r_1 r_2 )$ is the Gaussian curvature along the surface, $r_1$ and $r_2$ are two principal radii of curvature at any given location on the surface. $dC$ is a line element along the boundary formed by the cluster and solid and $\kappa_g$ is the geodesic curvature along the contact line $\partial M$. The simplest way to understand this equation is to study a sessile drop as displayed in Fig. \ref{fig1}(a).

We can use piece-wise integration to study the sessile drop. The surface manifold of the cluster $M$ can be partitioned into fluid/fluid interface $M_{ab}$, fluid/solid interface $M_{as}$ and a contact line $\partial M$. Average Gaussian curvature of each interface are $\kappa_{ab}$ and $\kappa_{as}$, respectively. The geodesic curvature along $\partial M$ is defined from two reference plans, i.e. the fluid/fluid interface ($\kappa_{g_{ab}}$) and fluid/solid interface ($\kappa_{g_{as}}$). The latter is the contact angle that is measured during a sessile drop experiment. The former is the angle over which the contact line deviates from being straight. This is commonly used to determine line tension, which is often disregarded in porous media with micro-meter sized pores. This geodesic curvature term corresponds to a total angle of change that is based on the total Gaussian curvature of the cluster as a consequence of its topology. From the aforementioned definitions, we arrive at the following formula for a 3D cluster,

\begin{align}
4\pi\chi(C)=\int_{M_{ab}} \kappa_{T}dS+\int_{M_{as}} \kappa_{T}dS+\int_{\partial M}(\kappa_{g_{ab}}+\kappa_{g_{as}})dC.
    \label{GBT-complex}	
\end{align}

The total Gaussian curvatures $\kappa_{T}$ of the two surface components appearing in Eq. (\ref{GBT-complex}) can be represented as the corresponding average curvature and surface area,

\begin{align}
\int_{M_{ab}} \kappa_{T}dS &= \kappa_{ab}  A_{ab},\\
\int_{M_{as}} \kappa_{T}dS &= \kappa_{as}  A_{as}.
\end{align}
By substituting above equations into Eq. (\ref{GBT-complex}), we obtain,

\begin{align}
4\pi\chi(C)=\kappa_{ab}  A_{ab}+\kappa_{as}  A_{as}+(\kappa_{g_{ab}}+\kappa_{g_{as}})L.
    \label{GBT-simple}	
\end{align}
where $L$ is the length of contact line. The geodesic curvature term, $\int_{\partial M}(\kappa_{g_{ab}}+\kappa_{g_{as}})dC$, can be represented by the deficit curvature, i.e. the summation of the geodesic curvature, and the length of contact line. Noting that the Euler characteristic for a single cluster is one, we therefore arrive at the following formula,

\begin{align}
\kappa_{g_{ab}} + \kappa_{g_{as}} = \frac{4 \pi - \kappa_{ab}A_{ab} - \kappa_{as}A_{as}}{L},
    \label{GBT-deficit}	
\end{align}
which explains the average curvature along the contact line for any geometrical arrangement of a fluid cluster where its topology is $\chi = 1$. This geometrical statement is true under any homeomorphic deformation. Examples of homeomorphic fluid droplets are provided in Fig. \ref{fig1}. By inspection, it is evident that the cluster in Figs. \ref{fig1}(a) and \ref{fig1}(b) both have the same Euler characteristic (topology). The implication, as stated in Eq. (\ref{GBT}), is that both fluid clusters must have the same total Gaussian curvature. Geometrically, the local curvature is not the same for these objects. Considering a cluster in 2D space, the total Gaussian curvature in Fig. \ref{fig1}(a) is distributed over the fluid/fluid interface and three-phase contact points. As a consequence, the total Gaussian curvature in the fluid/fluid interface is $2\pi(1-x)$ where $x$ is the fraction of the cluster surface area that is hypothetically extended into the solid substrate. The remaining curvature is in the contact points represented by the contact angle, $\theta$, since the fluid/solid interface is flat with zero curvature. For the cluster in real porous media as shown in Fig. \ref{fig1}(b), the multi-colored curves represent the cluster surface boundaries. The total Gaussian curvature in these interface can be determined by integrating over the surfaces. The remaining curvature would then be distributed over the contact points marked by red stars. 

\section{3. Materials and Methods}

\subsection{3.1. Experimental Set-up and Image Processing}
We performed a capillary pressure versus saturation experiment for Bentheimer sandstone. The sample was initially fully saturated with brine ($1.0$ $\rm M$ of KI to degassed DI water). Thereafter, brine was withdrawn from the sample at $18$ $\rm \mu L/hr$ through a syringe pump as air entered the core from the top. A semi-permeable hydrophilic membrane was placed at the bottom edge of the core to avoid air breakthrough from the core, which allowed for high air saturation to be achieved. The pump was sequentially stopped after a period of pumping and allowed $5$ $\rm min$ to equilibrate prior to measuring the pressure difference between the air and brine. Once irreducible brine saturation was reached, the process was reversed by injecting the brine phase. At irreducible gas saturation, the sample was imaged by using a bench-top helical micro-CT scanner at the Australian National University to visualize the 3D quasi-static spatial arrangement of the immiscible fluids. The core sample was imaged at $65$ $\rm amp$ and $100$ $\rm MeV$, and the scan took $1$ $\rm hr$ and $21$ $\rm min$ for a quasi-static 3D image at $4.95$ $\rm \mu m$ resolution. 

Image processing was then performed to obtain segmented 3D images of fluid clusters. Three-phase image segmentation was accomplished by performing two-phase segmentation (air and combined brine/solid phases) using an active contours segmentation routine \cite{sheppard2004techniques}, followed by registration and image arithmetic with the 'dry' scan of the sample (which contains only solid and air phases). Post-segmentation processes that including noise removal was achieved by removing speckles of air-identified voxels which were smaller than $1.4\times10^{\text{-}5}$ $\rm mm^3$, equivalent to a spherical pore radius of $15$ $\rm \mu m$. This allowed us to observe the spatial fluid configurations and measure the geometrical properties of each cluster in our tested porous medium at a resolution of $4.95$ $\rm \mu m$.

\subsection{3.2. Model and Simulation Set-up}
To test a range of wetting conditions, we use the multi-mineral model presented in \cite{shabaninejad2017pore}. It was generated from micro-CT imaging of a North Shore sandstone that was quantified in terms of its mineralogical content. The data set was segmented with each solid voxel labelled as as either quartz, clay or other. Further, we use the two-phase flow Lattice Boltzmann method (LBM) to simulate drainage and imbibition processes \cite{mcclure2014novel}. Details of the LBM method tested and applied in various porous media are presented elsewhere \cite{mcclure2014novel,armstrong2016beyond,armstrong2017flow,liu2018influence,mcclure2018geometric}. Two wetting states of the model are tested, one with an overall wettability of $0.1$ (neutral-wet) and the other with an overall wettability of $0.7$ (strongly water-wet). The overall wettability, $W$, is defined as the summation of the cosine of the contact angles assigned to the mineral surfaces,

\begin{equation}
W = \sum \frac{\sigma_{ai}-\sigma_{bi}}{\sigma_{ab}}\phi_i
\label{sum}
\end{equation}
where $\sigma$ is interfacial tension with subscripts for each fluid or mineral pair $(i)$ and $\phi_i$ is the solid fraction. 

For $W = 0.7$, we generated a homogeneous wetting system where wettability is constant for all mineral phases. For $W = 0.1$, we generated a heterogeneous wetting system based on the mineral type, but with water-wet conditions applied in the corners. This case is designed to mimic a common assumed condition for digital rock simulations and/or resorted state cores that have been aged in crude oil at connate water saturation \cite{kovscek1993pore,donaldson1969wettability}. 

All simulations were initiated from a morphological-based (maximum inscribed spheres) drainage state with $S_w=0.2$. Water flooding was simulated until the production curves starts to level off. Once the curves level off at $\rm 1 M$ time steps, a second set of simulations for secondary drainage were conducted starting from the end-point of the water flood. These simulations provided the hysteretic drainage and imbibition curves required to calculate the USBM index \cite{donaldson1969wettability}. Furthermore, phase distributions were used to measure contact angles and interfacial curvatures. For direct contact angle measurements, we applied the method as explained by \cite{alratrout2017automatic}. For curvature measurements, we used the approach as explained by \cite{armstrong2012linking}.

\section{4. Results and Discussions}
\begin{figure}
\centering\includegraphics[width=0.38\textwidth]{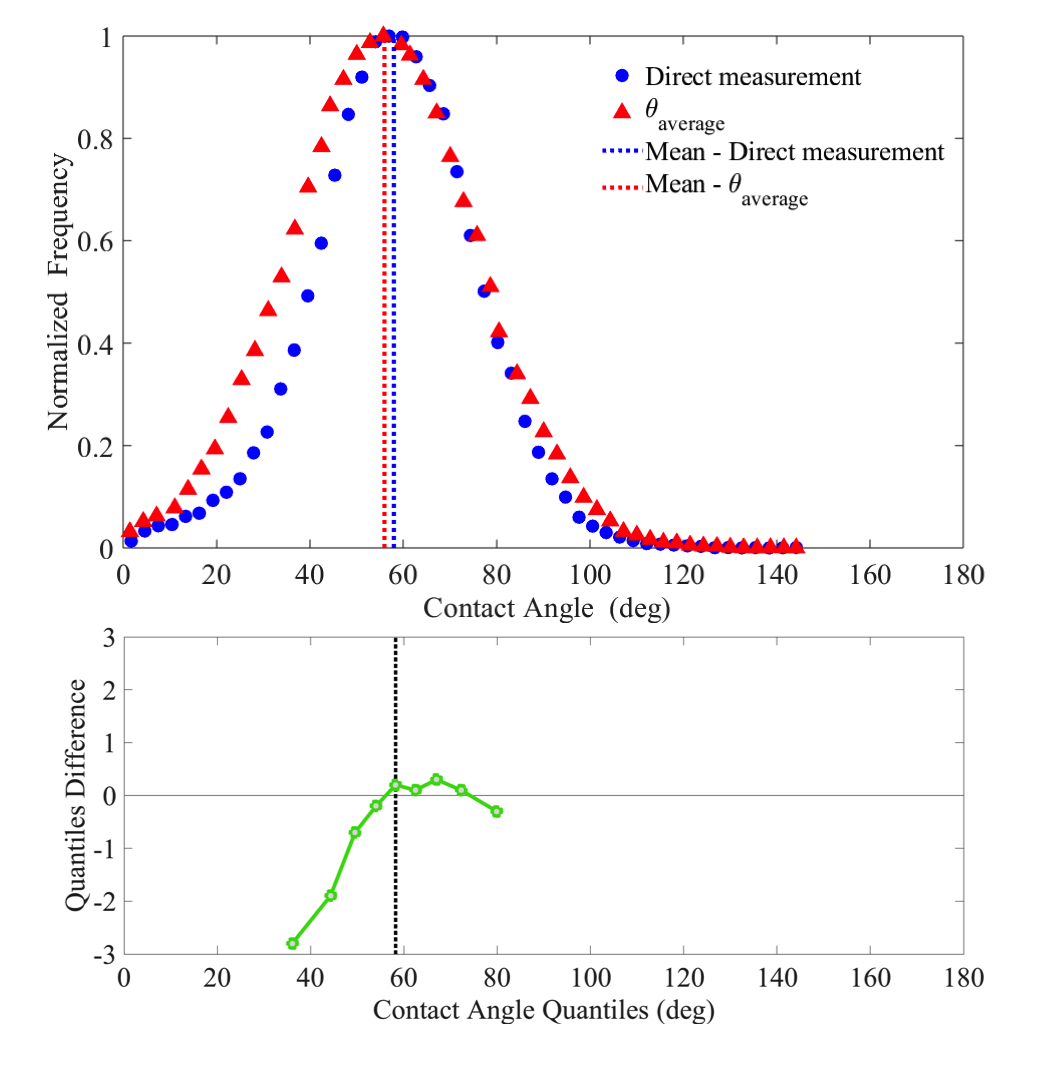}
\caption{A comparison between $\theta_{average}$ and direct pore-scale contact angle measurements. The mean values are within $2$-$3\%$ difference of each other.}
\label{fig3}
\end{figure}

Based on our proposed theory, Eq. (\ref{GBT-deficit}) can be constructed for a system of fluid clusters. As defined in Eq. (\ref{GBT-deficit}), $\kappa_{g_{as}} + \kappa_{g_{ab}}$ is the total deficit curvature for a cluster that must exist along the contact line. By considering surface and interfacial energies, it can be shown that $\kappa_{g_{as}} + \kappa_{g_{ab}} \approx \theta_{average}$ when $\kappa_{g_{as}} >> \kappa_{g_{ab}}$, where $\theta_{average}$ is the average contact angle measured for the cluster along the contact line. Combining these statements and Eq. (\ref{GBT-deficit}), we arrive at a formulation for a single cluster, 

\begin{align}
\theta_{average} = \frac{4 \pi - \kappa_{ab} A_{ab} -\kappa_{as}A_{as}}{L}.
\label{theta_average}	
\end{align}
To demonstrate and validate the proposed formulation, we verify the result by comparing contact angle measurements on segmented micro-CT images using Eq. (\ref{theta_average}) with contact angle measurements using a direct measurement method, as reported in \cite{alratrout2017automatic}. The results as shown in Fig. \ref{fig3} demonstrate that the difference is within a small quantile difference of only $2$-$3\%$ for the measured mean contact angle. The largest quantile difference is around $6\%$ for smaller contact angles. 

\subsection{4.1. Experimental Data Analysis}
\begin{figure}
\hspace*{-0.2cm}\includegraphics[width=0.5\textwidth]{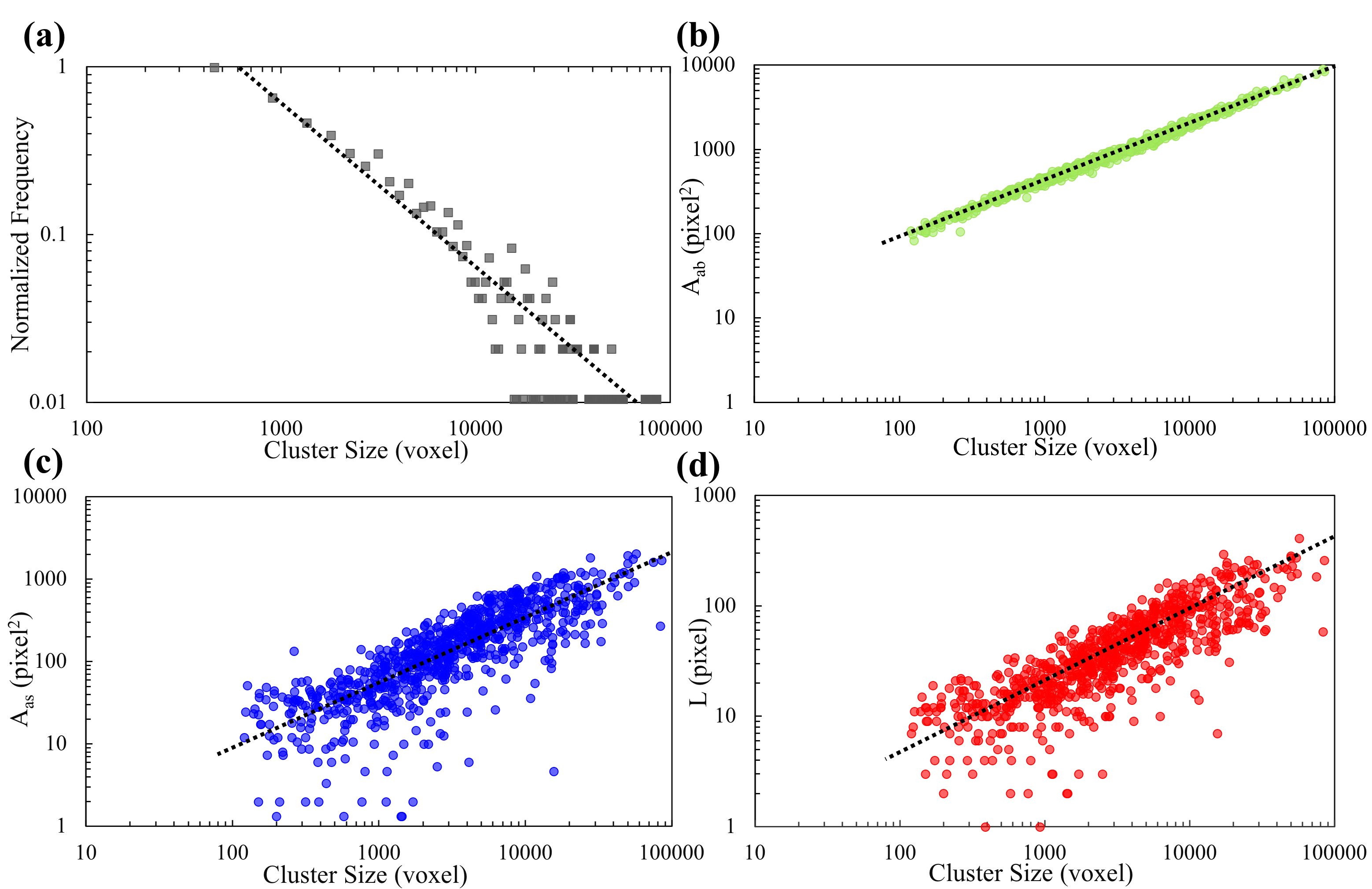}
\caption{(a) Cluster size distribution for sandstone experimental data. (b) Plot of fluid/fluid interfacial area $A_{ab}$ versus cluster size for the segmented image. (c) Plot of surface area between cluster and solid $A_{as}$ versus cluster size for the segmented image. (d) Plot of contact line length $L$ versus cluster size for the segmented image.}
\label{fig4}
\end{figure}

A commonly measured relationship for porous media is the cluster size distribution \cite{georgiadis2013pore}. The distribution is known to follow a power law relationship \cite{wilkinson1986percolation}. In Fig. \ref{fig4}(a), we provide the size distribution of fluid clusters measured for the sandstone sample. The distribution follows a typical distribution expected for systems that are near criticality. For fluid clusters near a percolation threshold, it is shown in Appendix A, a significant fraction of these clusters exist as $\chi(C)=1$. We also demonstrate, in Appendix B, that the average Gaussian curvature of the pore space ($\kappa_s$) can be used to represent $\kappa_{as}$. To obtain $\kappa_{s}$, we perform a representative elementary volume (REV) analysis using the segmented image. A representative value of $0.017 \mu m^{-2}$ was observed for the system. In addition, Figs. \ref{fig4}(b), \ref{fig4}(c) and \ref{fig4}(d) provide the relationship for $A_{ab}$, $A_{as}$, and $L$ versus cluster volume size. We observe that there is a positive correlation for $A_{ab}$, $A_{as}$ and $L$ versus cluster size that follows a linear relationship. 

\begin{figure}
\hspace*{-0.2cm}\includegraphics[width=0.5\textwidth]{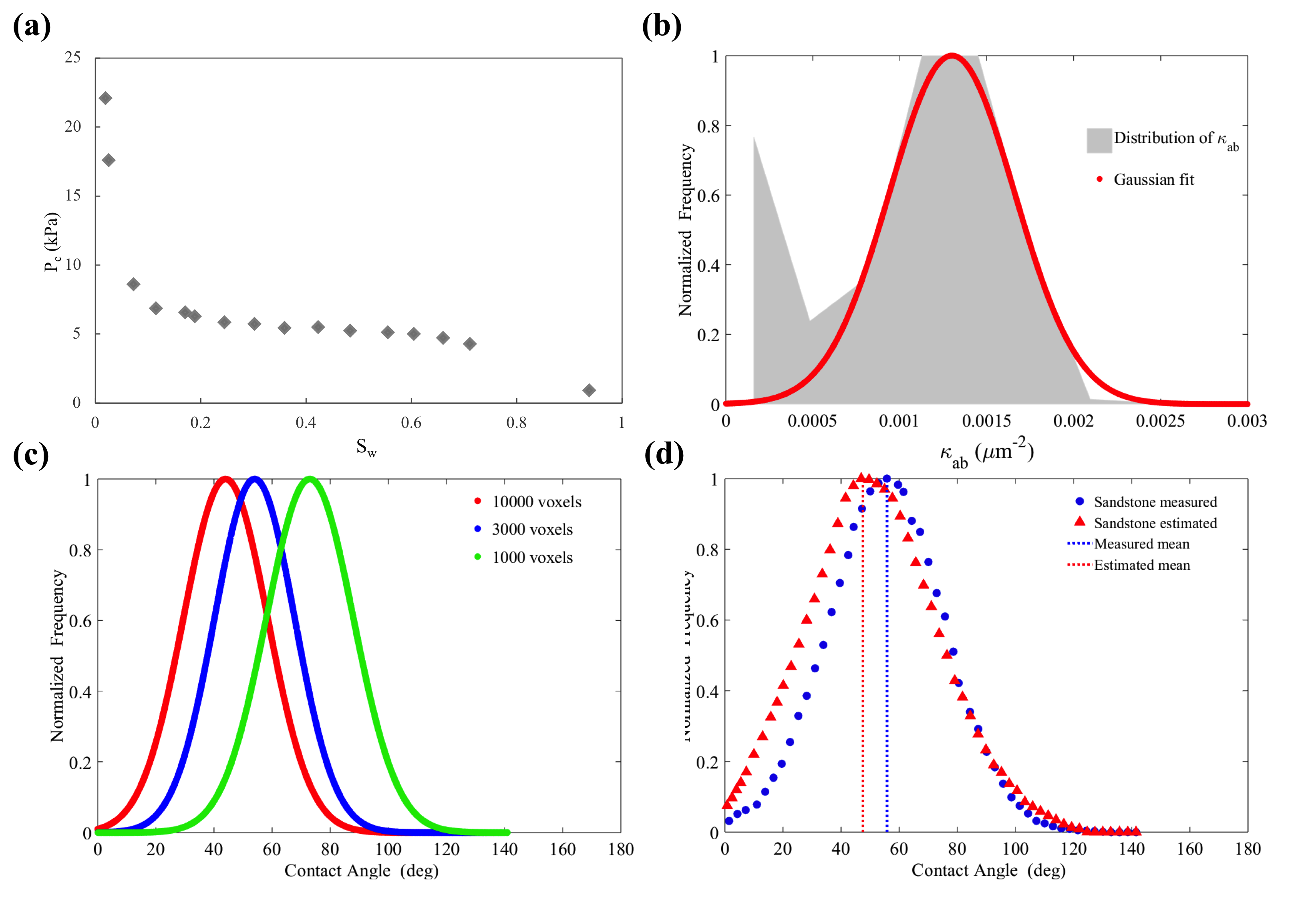}
\caption{(a) Capillary pressure saturation curve from the primary drainage experiment. (b) Normalized fluid/fluid Gaussian curvature distribution derived from the capillary pressure saturation curve with its Gaussian fit. (c) Normalized Gaussian distribution of contact angles for the selected three different cluster volumes based on petrophysical data. (d) Normalized distribution of contact angles for the whole porous medium measured by direct pore-scale methods compared to the developed probability-based approach. The result shows a similar mean contact angle for both approaches, which is $47.5^\circ$ and $55.8^\circ$, respectively.}
\label{fig5}
\end{figure}

To test the utility of Eq. (\ref{theta_average}), we attempt to predict $\theta_{average}$ by utilizing the relationship between capillary pressure and saturation. In Fig \ref{fig5}(a), we provide the capillary pressure versus saturation relationship for the tested porous medium. It provides insights into the physical structure or geometry of the system, which is known to represent the curvature of the fluid/fluid interfaces \cite{armstrong2012linking}. At equilibrium, capillary pressure reflects fluid/fluid interfacial curvature in terms of the Young-Laplace equation \cite{laplace1805traite}, 

\begin{align}
P_c = \sigma \big(\frac{1}{r_1} + \frac{1}{r_2}\big), 
\end{align}
where $\sigma$ denotes interfacial tension between the immiscible fluids. The two principals of curvature are $\kappa_1 = 1/r_1$ and $\kappa_2 = 1/r_2$, where $r_1$ and $r_2$ are the maximum and minimum radius of curvatures. Typical petrophysical approaches assume axisymmetric interfaces meaning that $r_1=r_2$ and then use $P_c(S_w)$ data to infer the size of the pore throat ($r$) that is entered by an invading fluid at a given saturation. Based on this assumption, the area under the $P_c(S_w)$ curve for a small range of $P_c$ provides the volume of the pore region that is filled when the capillary barrier of size $r$ is overcame. This information can be used to generate a distribution of characteristic length scales for a given porous medium, as depicted in Fig. \ref{fig5}(b). This distribution is reported as the probability distribution, $PDF(\kappa_{ab})$, for characteristic length scale, $1/r^2$, otherwise known as fluid/fluid interfacial Gaussian curvature. As will be shown, this approximation provides a reasonable first-order estimate for Gaussian curvature that provides accurate contact angle estimates for all of our tested systems. For porous media with fluid interfaces that are not axisymmetric, other means would be required to determine $PDF(\kappa_{ab})$. 

By taking a statistical mechanics approach whereby a macroscopic parameter can be represented by microscopic states that follows a given probability density function \cite{brown2003statistical}, we will attempt to predict the distribution of $\theta_{average}$ for an ensemble of clusters. We propose that a cluster of a given volume could be located anywhere in the porous medium based on the $PDF(\kappa_{ab})$. Therefore, various contact angles and Gaussian curvatures are possible for a single cluster volume. In Figs. \ref{fig4}(b), \ref{fig4}(c) and \ref{fig4}(d), the linear correlations for cluster surface areas and contact line length versus cluster volume size are provided and will be utilized in Eq. (\ref{theta_average}) for each cluster size. By applying this probability-based approach, Eq. (\ref{theta_average}) provides a distribution of possible $\theta_{average}$ for a cluster of a given volume. The $\theta_{average}$ distributions for three selected cluster volumes of $1000$, $3000$ and $10000$ voxels are shown in Fig. \ref{fig5}(c). The distributions predict that larger clusters tend to have smaller contact angles based on the geometrical constraints of Eq. (\ref{theta_average}). To verify this, in Table \ref{tab:Average}, we summarize the average contact angle values of $20$ clusters measured directly from the segmented image. It is evident that a similar trend can be demonstrated in the experimental data. On average the smaller clusters tend to have higher contact angles, given that smaller clusters also tend to have lower interfacial curvature. A similar tend between interfacial curvature and contact angle was reported by \cite{alratrout2018spatial}.
\begin{table}[htbp]
   \centering
   \caption{Summary of average contact angles for different cluster size.}
   \begin{tabular}{L{3cm}C{1.5cm}C{1.5cm}} 
      \hline 
      \hline
      Cluster Size (voxel) & n & $\overline \theta$\\
      \hline
      1000 & 20 & $76.5^\circ$\\
      5000 & 20 & $49.8^\circ$\\
      10000 & 20 & $43.6^\circ$\\
    \hline
    \hline
    \end{tabular}
   \label{tab:Average}
\end{table}

To determine the final distribution of contact angles for a system, we need to consider the volume distribution of the clusters. In Fig. \ref{fig4}(d), we combine the distributions of $\theta_{average}$ for each cluster volume using a weighted function based on cluster size distribution,
\begin{equation}
x_T(i) = \sum w_{j}x_{j}(i),
\label{sum_cluster}
\end{equation}
where $x(i)$ is the counts for each contact angle $i$, and $w_j$ is the weight for each cluster size $j$. Here $w(j) = N_j/N_T$, where $N_j$ is the number of clusters for size $j$ and $N_T$ is the total number of clusters. 

The result of Eq. (\ref{theta_average}) provides a distribution of $\theta_{average}$ that are comparable with those measured directly from pore-scale images, as provided in Fig. \ref{fig5}(d). The mean value using the probability-based approach versus direct measurement of contact angles from segmented data are $47.5^\circ$ and $55.8^\circ$, respectively. The difference in these values could result from non-axisymmetric interfaces. In addition, it could be noted that $P_c(S_w)$ provides higher curvature interfaces than what existed. It is common for interfaces to relax to lower curvature states. Since curvatures predicted by $P_c(S_w)$ relationship are related to pore throat sizes and not the larger pore bodies, it is likely that these curvatures are slightly greater than that observed during the experiment. As a consequence, the corresponding contact angles predicted are relatively smaller than that measured by direct methods. For instance, consider a typical cluster size of $3405$ voxels, which has a fluid/fluid Gaussian curvature of $0.00186 \mu m^{-2}$, the resulting contact angle from Eq. (\ref{theta_average}) is $56 ^\circ$. If we assign a larger curvature value of $0.002 \mu m^{-2}$, an angle of $53^\circ$ could be obtained.

\subsection{4.2. Simulation Data Analysis}
\begin{figure}
\hspace*{-0.1cm}\includegraphics[width=0.5\textwidth]{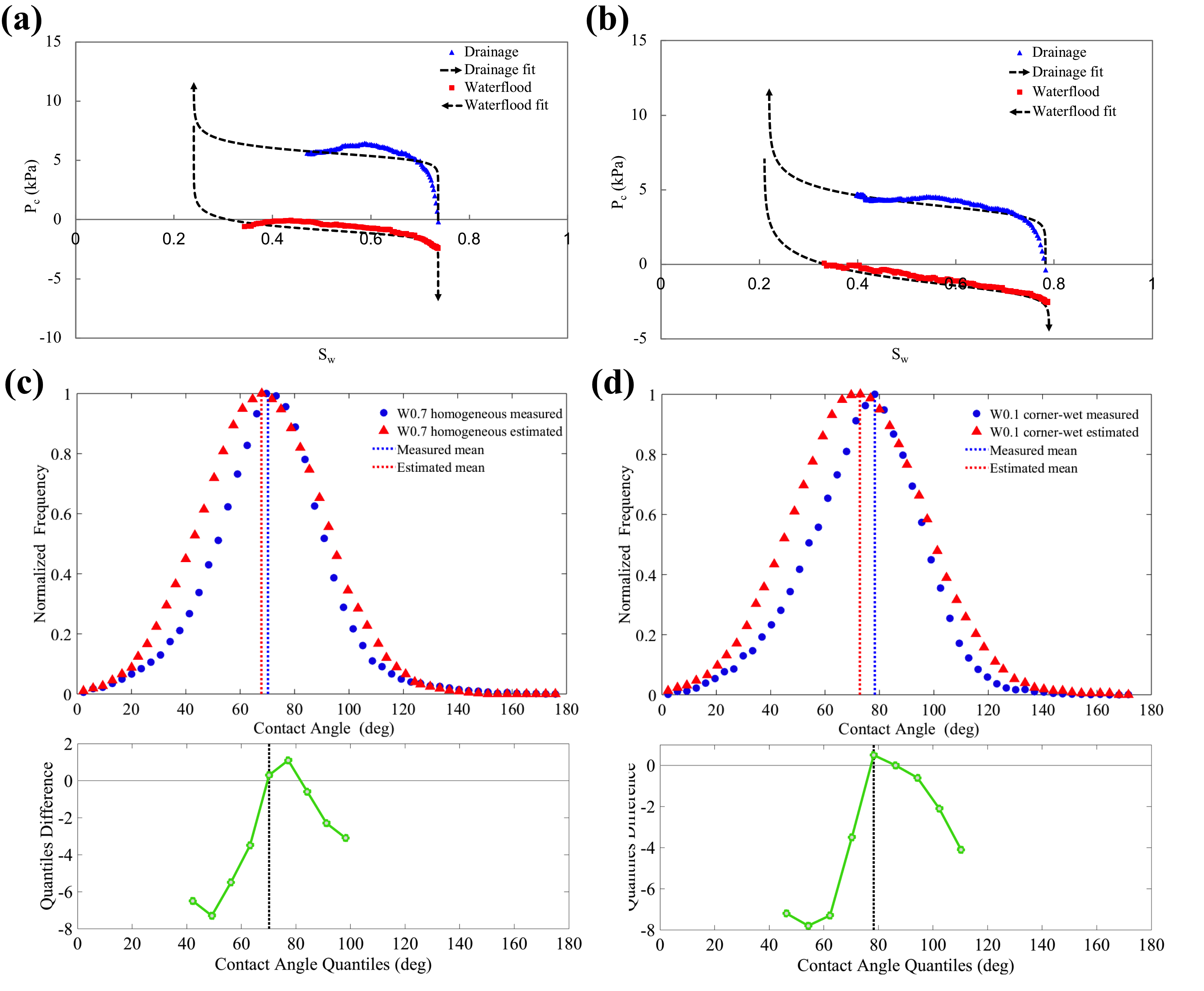}
\caption{Simulated capillary pressure versus saturation data for (a) $W=0.7$ homogeneous-wet case and (b) $W=0.1$ corner-wet case. Regression analysis is applied to fit Van Genuchten equation to the simulated data. Measured pore-scale contact angles compared to those predicted by Eq. (\ref{theta_average}) using $P_c(S_w)$ data for (c) $W=0.7$ homogeneous-wet case and (d) $W=0.1$ corner-wet case with percentage differences for each quantile of the contact angle distribution.}
\label{fig6}
\end{figure}

To further investigate our approach, we test simulation data in a multi-mineral model under various wetting conditions. The predicted and measured contact angle distributions are presented in Fig. \ref{fig6}. Both simulation cases result in a broad distribution of contact angles within a small quantile difference compared to direct measurements. The largest percentage difference results for the lower contact angle measurements with percentage differences around $4$-$8\%$. The mean values provide percentage differences of only $2$-$4\%$. It is interesting that the simulation cases provide a wide distribution of contact angles, as predicted by our statistical mechanics approach whereby all possible microscopic states of a fluid cluster are considered. 

In particular, the homogeneous case ($W=0.7$) does not result in a unique contact angle with a standard deviation of zero. Indeed, for the homogeneous case, the cosine of the equilibrium contact angles assigned to each mineral voxel is the same suggesting that the measured contact angle should be constant. However, this is not the result for various reasons. Contact angle hysteresis due to surface roughness, chemical heterogeneity and other flow dynamic effects must be considered when interpreting the data. In addition, how long it takes an interface to return to its equilibrium condition after an advancing or receding displacement is an open question \cite{schaffer2000contact,dye2016adaptive,schluter2017time}. For an imbibition process, there are complex sequences of cooperative dynamics at the pore-scale where interfaces are advancing and receding, providing a range of various contact angles. The presented distributions represent advancing and receding contact angles in addition to local regions were an interface could be pinned, such as the entrance to a pore body. Distributions, such as these have been reported for experimental data in carbonate rocks where surface chemistry is nearly homogeneous \cite{andrew2014pore}. Our geometrical approach captures these aspects since it considers all possible geometrical states for fluid clusters regardless of the underlying dynamics.

With the proposed geometrical approach states are largely defined by $P_c(S_w)$ relationship, a broader distribution of possible geometrical states results in a broader distribution of contact angles. This can be observed by comparing $P_c(S_w)$ curve along with the data fit to Van Genuchten equation in Figs. \ref{fig6}(a) and \ref{fig6}(b) to the resulting contact angle distributions in Figs. \ref{fig6}(c) and \ref{fig6}(d). The broader distribution of contact angles for corner-wet system is represented by the broader distribution of curvatures represented by $P_c(S_w)$ data. $P_c(S_w)$ curve for the $W=0.7$ homogeneous-wet case is flatter than that for $W=0.1$ corner-wet case, suggesting that there is a broader range of possible interfacial curvatures for the $W=0.1$ corner-wet case. $P_c(S_w)$ curve is also used to measure the USBM wettability indices for these two cases. Drainage and imbibition curves used to determine the USBM indices are provided in Figs. \ref{fig6}(a) and \ref{fig6}(b). In addition, a summary of pore-scale contact angle measurements and USBM wettability indices are provided in Table \ref{tab:summary}. 
\begin{table}[htbp]
   \centering
   \caption{Summary of pore-scale contact angle measurements and USBM wettability indices.}
   \begin{tabular}{lcccc} 
      \hline 
      \hline
      Case & W & USBM & Mean $\theta$ & Standard Deviation\\
      \hline
      Homogeneous-wet & $0.7$ & $1$ & $70.2^\circ$ & $20.5^\circ$\\
      Corner-wet & $0.1$ & $0.1$ & $78.3^\circ$ & $22.4^\circ$\\
    \hline
    \hline
    \end{tabular}
   \label{tab:summary}
\end{table}

It is interesting to note that a USBM index of $1$ results in a mean contact angle of $70.2^\circ$. It would generally be expected that a lower contact angle would be measured for such a water-wet USBM index. However, the measured contact angles are only a sampling of the grain surfaces where the contact line exists. The finding is a statement as to where on the grain surface the fluid/fluid interface prefers to be located under the prevailing conditions. The contact line is likely to be pinned at rough surfaces and/or sharp bends, resulting in contact angles greater than the intrinsic contact angle measured for a smooth and flat surface at equilibrium conditions. For corner-wet case, we measure a USBM index of $0.1$ with a mean contact angle of $78.3^\circ$. It represents a more neural-wet condition with a broader range of measured contact angels due to the broader range of surface energies applied to the modeling domain. We also measure a few contact angles that would represent water-wet conditions and these likely result from interfaces being pinned in corner regions that transition from water-wet to oil-wet states.

\section{5. Conclusion}
By utilizing $P_c(S_w)$ data, a distribution of contact angles for a collection of fluid clusters can be predicted by using the proposed formulation. It depends on the system having critical cluster-like behaviour and fluid interfaces that are axisymmetric. These assumptions are shown to provide reasonable results for the tested granular porous systems. For interface geometries that are anisotropic or saturation where both fluids are percolating, additional topological arguments would be necessary. While the universal relationship between topology and total Gaussian curvature of an object would still hold, i.e. Eq. (\ref{GBT}), the topology of the fluid phases would need to be directly measured and the current assumptions would need to be reassessed. 

The presented contact angle predictions are within $4 \%$ difference for mean contact angle and around $8 \%$ difference for smaller contact angles when compared with direct methods. However, it should be noted that direct pore-scale contact angle measurements also have error and in particular these errors would be more prevalent for smaller contact angles on rough and variable surfaces. Thus, the assumed benchmark should also be questioned. Despite these issues, the results are within a reasonable margin of difference and the developed framework provides new insights on the geometrical state of fluids and a link between commonly used pore-scale and continuum-scale metrics. By utilizing the proposed framework, we provide a reasonable distribution of pore-scale contact angles without the need for direct measurements. The proposed method provides a broad distribution due to contact angle hysteresis. The results are consistent with the cluster size distribution and $P_c(S_w)$ curve, which covers the up-scaled contact angle information. These distributions that are macroscopic measure could be more suitable for pore network models, direct numerical simulations, and development of more advanced theories for multiphase flow in porous media. 

\section{Acknowledgments}
\begin{acknowledgments}
C. S. acknowledges an Australian Government Research Training Program Scholarship. A. H. acknowledges ARC DE180100082 and the ANU/UNSW Digicore Research Consortium. J. M. acknowledges an award of computer time provided by the Department of Energy Early Science program. This research also used resources of the Oak Ridge Leadership Computing Facility, which is a DOE Office of Science User Facility supported under Contract DE-AC05-00OR22725.
\end{acknowledgments}
\appendix
\section{Appendix A: Supporting Material - Cluster Euler Characteristic Analysis}
We analyze the topology of each fluid cluster in sandstone data. It is found that a significant fraction of the clusters exists as $\chi(C)=1$. When clusters are formed in our experiment, we found that $99$ $\%$ of the clusters have a measured Euler characteristic of one. This result demonstrates that Eq. (\ref{GBT-deficit}) is applicable for our fluid system since all clusters are homeomorphic to a sessile drop. For a system where phase saturation is increased, Eq. (\ref{GBT-deficit}) would have to be developed for each phase topology of increasing connectivity. The equation can simply be developed for a torus, double torus and so forth to provide homeomorphic objects for each class of cluster. 

\section{Appendix B: Supporting Material - Representative Elementary Volume Analysis of $\kappa_{s}$}
\begin{figure}[h!]
\includegraphics[width=0.51\textwidth]{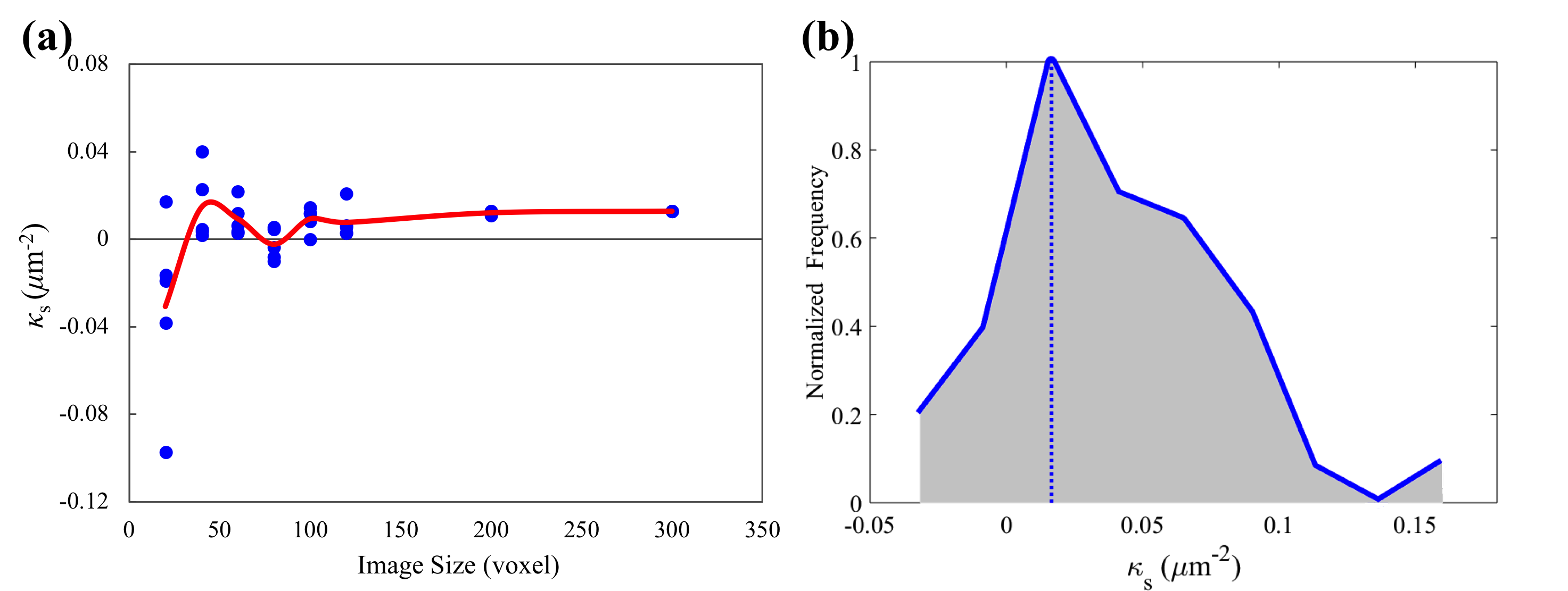}
\caption{(a) REV analysis of Gaussian curvature for solid surface ($\kappa_{s}$). The average $\kappa_s$ value comes to $0.017$ $\mu m^{-2}$. (b) Distribution of $\kappa_{s}$ are measured directly from segmented multiphase image.}
\label{REV_Gaussian}
\end{figure}

We perform a representative elementary volume (REV) analysis using segmented image of tested sandstone sample to obtain an average $\kappa_s$. It is important that the Gaussian curvature of a sub-volume of the porous medium is representative of the whole porous structure at a larger scale. In this study, we randomly choose the location of sub-volumes in the image for sizes $20^3$, $40^3$, $60^3$, $80^3$, $100^3$, $120^3$, $200^3$ and $300^3$ voxels. The results indicate that an image size larger than $200^3$ voxels can be considered representative for Gaussian curvature. The average $\kappa_s$ value obtained is $0.017$ $\mu m^{-2}$ as shown in Fig. \ref{REV_Gaussian}(a). In addition, in Fig. \ref{REV_Gaussian}(b), Gaussian curvature of fluid/solid interfaces are also measured directly from segmented multiphase image and the average measured value agrees with that obtained from REV analysis. These results suggest that the average Gaussian curvature of sandstone surface is a reasonable estimate for $\kappa_s$, i.e., the average Gaussian curvature for fluid/solid interface for a collection of clusters. 


\begin{thebibliography}{40}%
\makeatletter
\providecommand \@ifxundefined [1]{%
 \@ifx{#1\undefined}
}%
\providecommand \@ifnum [1]{%
 \ifnum #1\expandafter \@firstoftwo
 \else \expandafter \@secondoftwo
 \fi
}%
\providecommand \@ifx [1]{%
 \ifx #1\expandafter \@firstoftwo
 \else \expandafter \@secondoftwo
 \fi
}%
\providecommand \natexlab [1]{#1}%
\providecommand \enquote  [1]{``#1''}%
\providecommand \bibnamefont  [1]{#1}%
\providecommand \bibfnamefont [1]{#1}%
\providecommand \citenamefont [1]{#1}%
\providecommand \href@noop [0]{\@secondoftwo}%
\providecommand \href [0]{\begingroup \@sanitize@url \@href}%
\providecommand \@href[1]{\@@startlink{#1}\@@href}%
\providecommand \@@href[1]{\endgroup#1\@@endlink}%
\providecommand \@sanitize@url [0]{\catcode `\\12\catcode `\$12\catcode
  `\&12\catcode `\#12\catcode `\^12\catcode `\_12\catcode `\%12\relax}%
\providecommand \@@startlink[1]{}%
\providecommand \@@endlink[0]{}%
\providecommand \url  [0]{\begingroup\@sanitize@url \@url }%
\providecommand \@url [1]{\endgroup\@href {#1}{\urlprefix }}%
\providecommand \urlprefix  [0]{URL }%
\providecommand \Eprint [0]{\href }%
\providecommand \doibase [0]{https://doi.org/}%
\providecommand \selectlanguage [0]{\@gobble}%
\providecommand \bibinfo  [0]{\@secondoftwo}%
\providecommand \bibfield  [0]{\@secondoftwo}%
\providecommand \translation [1]{[#1]}%
\providecommand \BibitemOpen [0]{}%
\providecommand \bibitemStop [0]{}%
\providecommand \bibitemNoStop [0]{.\EOS\space}%
\providecommand \EOS [0]{\spacefactor3000\relax}%
\providecommand \BibitemShut  [1]{\csname bibitem#1\endcsname}%
\let\auto@bib@innerbib\@empty
\bibitem [{\citenamefont {Bonn}\ \emph {et~al.}(2009)\citenamefont {Bonn},
  \citenamefont {Eggers}, \citenamefont {Indekeu}, \citenamefont {Meunier},\
  and\ \citenamefont {Rolley}}]{bonn2009wetting}%
  \BibitemOpen
  \bibfield  {author} {\bibinfo {author} {\bibfnamefont {D.}~\bibnamefont
  {Bonn}}, \bibinfo {author} {\bibfnamefont {J.}~\bibnamefont {Eggers}},
  \bibinfo {author} {\bibfnamefont {J.}~\bibnamefont {Indekeu}}, \bibinfo
  {author} {\bibfnamefont {J.}~\bibnamefont {Meunier}}, and\ \bibinfo {author}
  {\bibfnamefont {E.}~\bibnamefont {Rolley}},\ }\href
  {https://doi.org/10.1103/RevModPhys.81.739} {\bibfield  {journal} {\bibinfo
  {journal} {Reviews of modern physics}\ }\textbf {\bibinfo {volume} {81}},\
  \bibinfo {pages} {739} (\bibinfo {year} {2009})}\BibitemShut {NoStop}%
\bibitem [{\citenamefont {De~Gennes}(1985)}]{de1985wetting}%
  \BibitemOpen
  \bibfield  {author} {\bibinfo {author} {\bibfnamefont {P.-G.}\ \bibnamefont
  {De~Gennes}},\ }\href {https://doi.org/10.1103/RevModPhys.57.827} {\bibfield
  {journal} {\bibinfo  {journal} {Reviews of modern physics}\ }\textbf
  {\bibinfo {volume} {57}},\ \bibinfo {pages} {827} (\bibinfo {year}
  {1985})}\BibitemShut {NoStop}%
\bibitem [{\citenamefont {Weitz}(2007)}]{Weitz2007Dripping}%
  \BibitemOpen
  \bibfield  {author} {\bibinfo {author} {\bibfnamefont {D.}~\bibnamefont
  {Weitz}},\ }\href@noop {} {\bibfield  {journal} {\bibinfo  {journal} {Mrs
  Bulletin}\ }\textbf {\bibinfo {volume} {32}},\ \bibinfo {pages} {702}
  (\bibinfo {year} {2007})}\BibitemShut {NoStop}%
\bibitem [{\citenamefont {Bauters}\ \emph {et~al.}(2000)\citenamefont
  {Bauters}, \citenamefont {Steenhuis}, \citenamefont {DiCarlo}, \citenamefont
  {Nieber}, \citenamefont {Dekker}, \citenamefont {Ritsema}, \citenamefont
  {Parlange},\ and\ \citenamefont {Haverkamp}}]{bauters2000physics}%
  \BibitemOpen
  \bibfield  {author} {\bibinfo {author} {\bibfnamefont {T.}~\bibnamefont
  {Bauters}}, \bibinfo {author} {\bibfnamefont {T.}~\bibnamefont {Steenhuis}},
  \bibinfo {author} {\bibfnamefont {D.}~\bibnamefont {DiCarlo}}, \bibinfo
  {author} {\bibfnamefont {J.~L.}\ \bibnamefont {Nieber}}, \bibinfo {author}
  {\bibfnamefont {L.}~\bibnamefont {Dekker}}, \bibinfo {author} {\bibfnamefont
  {C.}~\bibnamefont {Ritsema}}, \bibinfo {author} {\bibfnamefont {J.-Y.}\
  \bibnamefont {Parlange}}, and\ \bibinfo {author} {\bibfnamefont
  {R.}~\bibnamefont {Haverkamp}},\ }\href
  {https://doi.org/10.1016/S0022-1694(00)00197-9} {\bibfield  {journal}
  {\bibinfo  {journal} {Journal of Hydrology}\ }\textbf {\bibinfo {volume}
  {231}},\ \bibinfo {pages} {233} (\bibinfo {year} {2000})}\BibitemShut
  {NoStop}%
\bibitem [{\citenamefont {Morrow}(1970)}]{morrow1970physics}%
  \BibitemOpen
  \bibfield  {author} {\bibinfo {author} {\bibfnamefont {N.~R.}\ \bibnamefont
  {Morrow}},\ }\href {http://doi.org/10.1021/ie50726a006} {\bibfield  {journal}
  {\bibinfo  {journal} {Industrial \& Engineering Chemistry}\ }\textbf
  {\bibinfo {volume} {62}},\ \bibinfo {pages} {32} (\bibinfo {year}
  {1970})}\BibitemShut {NoStop}%
\bibitem [{\citenamefont {Botto}\ \emph {et~al.}(2017)\citenamefont {Botto},
  \citenamefont {Fuchs}, \citenamefont {Fouke}, \citenamefont {Clarens},
  \citenamefont {Freiburg}, \citenamefont {Berger},\ and\ \citenamefont
  {Werth}}]{botto2017effects}%
  \BibitemOpen
  \bibfield  {author} {\bibinfo {author} {\bibfnamefont {J.}~\bibnamefont
  {Botto}}, \bibinfo {author} {\bibfnamefont {S.~J.}\ \bibnamefont {Fuchs}},
  \bibinfo {author} {\bibfnamefont {B.~W.}\ \bibnamefont {Fouke}}, \bibinfo
  {author} {\bibfnamefont {A.~F.}\ \bibnamefont {Clarens}}, \bibinfo {author}
  {\bibfnamefont {J.~T.}\ \bibnamefont {Freiburg}}, \bibinfo {author}
  {\bibfnamefont {P.~M.}\ \bibnamefont {Berger}}, and\ \bibinfo {author}
  {\bibfnamefont {C.~J.}\ \bibnamefont {Werth}},\ }\href@noop {} {\bibfield
  {journal} {\bibinfo  {journal} {Energy \& Fuels}\ }\textbf {\bibinfo {volume}
  {31}},\ \bibinfo {pages} {5275} (\bibinfo {year} {2017})}\BibitemShut
  {NoStop}%
\bibitem [{\citenamefont {PaulGatenholm}\ \emph {et~al.}(1990)\citenamefont
  {PaulGatenholm}, \citenamefont {ChrisBonnerup},\ and\ \citenamefont
  {EvaWallstrom}}]{PaulGatenholm1990Wetting}%
  \BibitemOpen
  \bibfield  {author} {\bibinfo {author} {\bibnamefont {PaulGatenholm}},
  \bibinfo {author} {\bibnamefont {ChrisBonnerup}}, and\ \bibinfo {author}
  {\bibnamefont {EvaWallstrom}},\ }\href@noop {} {\bibfield  {journal}
  {\bibinfo  {journal} {Journal of Adhesion Science \& Technology}\ }\textbf
  {\bibinfo {volume} {4}},\ \bibinfo {pages} {11} (\bibinfo {year}
  {1990})}\BibitemShut {NoStop}%
\bibitem [{\citenamefont {Blossey}(2003)}]{blossey2003self}%
  \BibitemOpen
  \bibfield  {author} {\bibinfo {author} {\bibfnamefont {R.}~\bibnamefont
  {Blossey}},\ }\href {https://www.nature.com/articles/nmat856} {\bibfield
  {journal} {\bibinfo  {journal} {Nature materials}\ }\textbf {\bibinfo
  {volume} {2}},\ \bibinfo {pages} {301} (\bibinfo {year} {2003})}\BibitemShut
  {NoStop}%
\bibitem [{\citenamefont {Anderson}\ \emph {et~al.}(1987)\citenamefont
  {Anderson} \emph {et~al.}}]{anderson1987wettability}%
  \BibitemOpen
  \bibfield  {author} {\bibinfo {author} {\bibfnamefont {W.~G.}\ \bibnamefont
  {Anderson}} \emph {et~al.},\ }\href {https://doi.org/10.2118/16471-PA}
  {\bibfield  {journal} {\bibinfo  {journal} {Journal of petroleum technology}\
  }\textbf {\bibinfo {volume} {39}},\ \bibinfo {pages} {1} (\bibinfo {year}
  {1987})}\BibitemShut {NoStop}%
\bibitem [{\citenamefont {Anderson}\ \emph {et~al.}(1986)\citenamefont
  {Anderson} \emph {et~al.}}]{anderson1986wettability}%
  \BibitemOpen
  \bibfield  {author} {\bibinfo {author} {\bibfnamefont {W.}~\bibnamefont
  {Anderson}} \emph {et~al.},\ }\href@noop {} {\bibfield  {journal} {\bibinfo
  {journal} {Journal of petroleum technology}\ }\textbf {\bibinfo {volume}
  {38}},\ \bibinfo {pages} {1} (\bibinfo {year} {1986})}\BibitemShut {NoStop}%
\bibitem [{\citenamefont {Kovscek}\ \emph {et~al.}(1993)\citenamefont
  {Kovscek}, \citenamefont {Wong},\ and\ \citenamefont
  {Radke}}]{kovscek1993pore}%
  \BibitemOpen
  \bibfield  {author} {\bibinfo {author} {\bibfnamefont {A.}~\bibnamefont
  {Kovscek}}, \bibinfo {author} {\bibfnamefont {H.}~\bibnamefont {Wong}}, and\
  \bibinfo {author} {\bibfnamefont {C.}~\bibnamefont {Radke}},\ }\href@noop {}
  {\bibfield  {journal} {\bibinfo  {journal} {AIChE Journal}\ }\textbf
  {\bibinfo {volume} {39}},\ \bibinfo {pages} {1072} (\bibinfo {year}
  {1993})}\BibitemShut {NoStop}%
\bibitem [{\citenamefont {Donaldson}\ \emph {et~al.}(1969)\citenamefont
  {Donaldson}, \citenamefont {Thomas}, \citenamefont {Lorenz} \emph
  {et~al.}}]{donaldson1969wettability}%
  \BibitemOpen
  \bibfield  {author} {\bibinfo {author} {\bibfnamefont {E.~C.}\ \bibnamefont
  {Donaldson}}, \bibinfo {author} {\bibfnamefont {R.~D.}\ \bibnamefont
  {Thomas}}, \bibinfo {author} {\bibfnamefont {P.~B.}\ \bibnamefont {Lorenz}},
  \emph {et~al.},\ }\href {https://doi.org/10.2118/2338-PA} {\bibfield
  {journal} {\bibinfo  {journal} {Society of Petroleum Engineers Journal}\
  }\textbf {\bibinfo {volume} {9}},\ \bibinfo {pages} {13} (\bibinfo {year}
  {1969})}\BibitemShut {NoStop}%
\bibitem [{\citenamefont {Prodanovic}\ \emph {et~al.}(2006)\citenamefont
  {Prodanovic}, \citenamefont {Lindquist},\ and\ \citenamefont
  {Seright}}]{Prodanovic2006}%
  \BibitemOpen
  \bibfield  {author} {\bibinfo {author} {\bibfnamefont {M.}~\bibnamefont
  {Prodanovic}}, \bibinfo {author} {\bibfnamefont {W.~B.}\ \bibnamefont
  {Lindquist}}, and\ \bibinfo {author} {\bibfnamefont {R.~S.}\ \bibnamefont
  {Seright}},\ }in\ \href
  {https://pdfs.semanticscholar.org/697c/8e17c30f6663247ec39784db2b55af588d55.pdf}
  {\emph {\bibinfo {booktitle} {Proceedings of the XVI International Conference
  on Computational Methods in Water Resources, Copenhagen, Denmark 18-22 June
  2006}}}\ (\bibinfo {year} {2006})\BibitemShut {NoStop}%
\bibitem [{\citenamefont {Andrew}\ \emph {et~al.}(2014)\citenamefont {Andrew},
  \citenamefont {Bijeljic},\ and\ \citenamefont {Blunt}}]{andrew2014pore}%
  \BibitemOpen
  \bibfield  {author} {\bibinfo {author} {\bibfnamefont {M.}~\bibnamefont
  {Andrew}}, \bibinfo {author} {\bibfnamefont {B.}~\bibnamefont {Bijeljic}},
  and\ \bibinfo {author} {\bibfnamefont {M.~J.}\ \bibnamefont {Blunt}},\ }\href
  {https://doi.org/10.1016/j.advwatres.2014.02.014} {\bibfield  {journal}
  {\bibinfo  {journal} {Advances in Water Resources}\ }\textbf {\bibinfo
  {volume} {68}},\ \bibinfo {pages} {24} (\bibinfo {year} {2014})}\BibitemShut
  {NoStop}%
\bibitem [{\citenamefont {AlRatrout}\ \emph {et~al.}(2017)\citenamefont
  {AlRatrout}, \citenamefont {Raeini}, \citenamefont {Bijeljic},\ and\
  \citenamefont {Blunt}}]{alratrout2017automatic}%
  \BibitemOpen
  \bibfield  {author} {\bibinfo {author} {\bibfnamefont {A.}~\bibnamefont
  {AlRatrout}}, \bibinfo {author} {\bibfnamefont {A.~Q.}\ \bibnamefont
  {Raeini}}, \bibinfo {author} {\bibfnamefont {B.}~\bibnamefont {Bijeljic}},
  and\ \bibinfo {author} {\bibfnamefont {M.~J.}\ \bibnamefont {Blunt}},\ }\href
  {https://doi.org/10.1016/j.advwatres.2017.07.018} {\bibfield  {journal}
  {\bibinfo  {journal} {Advances in water resources}\ }\textbf {\bibinfo
  {volume} {109}},\ \bibinfo {pages} {158} (\bibinfo {year}
  {2017})}\BibitemShut {NoStop}%
\bibitem [{\citenamefont {Scanziani}\ \emph {et~al.}(2017)\citenamefont
  {Scanziani}, \citenamefont {Singh}, \citenamefont {Blunt},\ and\
  \citenamefont {Guadagnini}}]{scanziani2017automatic}%
  \BibitemOpen
  \bibfield  {author} {\bibinfo {author} {\bibfnamefont {A.}~\bibnamefont
  {Scanziani}}, \bibinfo {author} {\bibfnamefont {K.}~\bibnamefont {Singh}},
  \bibinfo {author} {\bibfnamefont {M.~J.}\ \bibnamefont {Blunt}}, and\
  \bibinfo {author} {\bibfnamefont {A.}~\bibnamefont {Guadagnini}},\ }\href
  {https://doi.org/10.1016/j.jcis.2017.02.005} {\bibfield  {journal} {\bibinfo
  {journal} {Journal of colloid and interface science}\ }\textbf {\bibinfo
  {volume} {496}},\ \bibinfo {pages} {51} (\bibinfo {year} {2017})}\BibitemShut
  {NoStop}%
\bibitem [{\citenamefont {McPhee}\ \emph {et~al.}(2015)\citenamefont {McPhee},
  \citenamefont {Reed},\ and\ \citenamefont
  {Zubizarreta}}]{mcphee2015wettability}%
  \BibitemOpen
  \bibfield  {author} {\bibinfo {author} {\bibfnamefont {C.}~\bibnamefont
  {McPhee}}, \bibinfo {author} {\bibfnamefont {J.}~\bibnamefont {Reed}}, and\
  \bibinfo {author} {\bibfnamefont {I.}~\bibnamefont {Zubizarreta}},\ }in\
  \href@noop {} {\emph {\bibinfo {booktitle} {Developments in Petroleum
  Science}}},\ Vol.~\bibinfo {volume} {64}\ (\bibinfo  {publisher} {Elsevier},\
  \bibinfo {year} {2015})\ pp.\ \bibinfo {pages} {313--345}\BibitemShut
  {NoStop}%
\bibitem [{\citenamefont {Amott}\ \emph {et~al.}(1959)\citenamefont {Amott}
  \emph {et~al.}}]{amott1959observations}%
  \BibitemOpen
  \bibfield  {author} {\bibinfo {author} {\bibfnamefont {E.}~\bibnamefont
  {Amott}} \emph {et~al.},\ }\href
  {https://www.onepetro.org/general/SPE-1167-G} {\bibfield  {journal} {\bibinfo
   {journal} {Petroleum Transactions}\ } (\bibinfo {year} {1959})}\BibitemShut
  {NoStop}%
\bibitem [{\citenamefont {Young}\ \emph {et~al.}(1805)\citenamefont {Young}
  \emph {et~al.}}]{young1805iii}%
  \BibitemOpen
  \bibfield  {author} {\bibinfo {author} {\bibfnamefont {T.}~\bibnamefont
  {Young}} \emph {et~al.},\ }\href {https://doi.org/10.1098/rstl.1805.0005}
  {\bibfield  {journal} {\bibinfo  {journal} {Philosophical transactions of the
  royal society of London}\ }\textbf {\bibinfo {volume} {95}},\ \bibinfo
  {pages} {65} (\bibinfo {year} {1805})}\BibitemShut {NoStop}%
\bibitem [{\citenamefont {Morrow}\ \emph {et~al.}(1975)\citenamefont {Morrow}
  \emph {et~al.}}]{morrow1975effects}%
  \BibitemOpen
  \bibfield  {author} {\bibinfo {author} {\bibfnamefont {N.~R.}\ \bibnamefont
  {Morrow}} \emph {et~al.},\ }\href {https://doi.org/10.2118/75-04-04}
  {\bibfield  {journal} {\bibinfo  {journal} {Journal of Canadian Petroleum
  Technology}\ }\textbf {\bibinfo {volume} {14}} (\bibinfo {year}
  {1975})}\BibitemShut {NoStop}%
\bibitem [{\citenamefont {Cassie}\ and\ \citenamefont
  {Baxter}(1944)}]{cassie1944wettability}%
  \BibitemOpen
  \bibfield  {author} {\bibinfo {author} {\bibfnamefont {A.}~\bibnamefont
  {Cassie}}and\ \bibinfo {author} {\bibfnamefont {S.}~\bibnamefont {Baxter}},\
  }\href {http://doi.org/10.1039/TF9444000546} {\bibfield  {journal} {\bibinfo
  {journal} {Transactions of the Faraday society}\ }\textbf {\bibinfo {volume}
  {40}},\ \bibinfo {pages} {546} (\bibinfo {year} {1944})}\BibitemShut
  {NoStop}%
\bibitem [{\citenamefont {Indekeu}(1994)}]{indekeu1994line}%
  \BibitemOpen
  \bibfield  {author} {\bibinfo {author} {\bibfnamefont {J.}~\bibnamefont
  {Indekeu}},\ }\href {https://doi.org/10.1142/S0217979294000129} {\bibfield
  {journal} {\bibinfo  {journal} {International Journal of Modern Physics B}\
  }\textbf {\bibinfo {volume} {8}},\ \bibinfo {pages} {309} (\bibinfo {year}
  {1994})}\BibitemShut {NoStop}%
\bibitem [{\citenamefont {Heslot}\ \emph {et~al.}(1990)\citenamefont {Heslot},
  \citenamefont {Cazabat}, \citenamefont {Levinson},\ and\ \citenamefont
  {Fraysse}}]{heslot1990experiments}%
  \BibitemOpen
  \bibfield  {author} {\bibinfo {author} {\bibfnamefont {F.}~\bibnamefont
  {Heslot}}, \bibinfo {author} {\bibfnamefont {A.}~\bibnamefont {Cazabat}},
  \bibinfo {author} {\bibfnamefont {P.}~\bibnamefont {Levinson}}, and\ \bibinfo
  {author} {\bibfnamefont {N.}~\bibnamefont {Fraysse}},\ }\href
  {https://doi.org/10.1103/PhysRevLett.65.599} {\bibfield  {journal} {\bibinfo
  {journal} {Physical Review Letters}\ }\textbf {\bibinfo {volume} {65}},\
  \bibinfo {pages} {599} (\bibinfo {year} {1990})}\BibitemShut {NoStop}%
\bibitem [{\citenamefont {Sch{\"a}ffer}\ and\ \citenamefont
  {Wong}(2000)}]{schaffer2000contact}%
  \BibitemOpen
  \bibfield  {author} {\bibinfo {author} {\bibfnamefont {E.}~\bibnamefont
  {Sch{\"a}ffer}}and\ \bibinfo {author} {\bibfnamefont {P.-z.}\ \bibnamefont
  {Wong}},\ }\href@noop {} {\bibfield  {journal} {\bibinfo  {journal} {Physical
  Review E}\ }\textbf {\bibinfo {volume} {61}},\ \bibinfo {pages} {5257}
  (\bibinfo {year} {2000})}\BibitemShut {NoStop}%
\bibitem [{\citenamefont {Chern}(1944)}]{chern1944simple}%
  \BibitemOpen
  \bibfield  {author} {\bibinfo {author} {\bibfnamefont {S.-s.}\ \bibnamefont
  {Chern}},\ }\href {https://www.jstor.org/stable/1969302} {\bibfield
  {journal} {\bibinfo  {journal} {Annals of mathematics}\ ,\ \bibinfo {pages}
  {747}} (\bibinfo {year} {1944})}\BibitemShut {NoStop}%
\bibitem [{\citenamefont {Sheppard}\ \emph {et~al.}(2004)\citenamefont
  {Sheppard}, \citenamefont {Sok},\ and\ \citenamefont
  {Averdunk}}]{sheppard2004techniques}%
  \BibitemOpen
  \bibfield  {author} {\bibinfo {author} {\bibfnamefont {A.~P.}\ \bibnamefont
  {Sheppard}}, \bibinfo {author} {\bibfnamefont {R.~M.}\ \bibnamefont {Sok}},
  and\ \bibinfo {author} {\bibfnamefont {H.}~\bibnamefont {Averdunk}},\ }\href
  {https://doi.org/10.1016/j.physa.2004.03.057} {\bibfield  {journal} {\bibinfo
   {journal} {Physica A: Statistical mechanics and its applications}\ }\textbf
  {\bibinfo {volume} {339}},\ \bibinfo {pages} {145} (\bibinfo {year}
  {2004})}\BibitemShut {NoStop}%
\bibitem [{\citenamefont {Shabaninejad}\ \emph {et~al.}(2017)\citenamefont
  {Shabaninejad}, \citenamefont {Middleton}, \citenamefont {Latham},\ and\
  \citenamefont {Fogden}}]{shabaninejad2017pore}%
  \BibitemOpen
  \bibfield  {author} {\bibinfo {author} {\bibfnamefont {M.}~\bibnamefont
  {Shabaninejad}}, \bibinfo {author} {\bibfnamefont {J.}~\bibnamefont
  {Middleton}}, \bibinfo {author} {\bibfnamefont {S.}~\bibnamefont {Latham}},
  and\ \bibinfo {author} {\bibfnamefont {A.}~\bibnamefont {Fogden}},\
  }\href@noop {} {\bibfield  {journal} {\bibinfo  {journal} {Energy \& fuels}\
  }\textbf {\bibinfo {volume} {31}},\ \bibinfo {pages} {13221} (\bibinfo {year}
  {2017})}\BibitemShut {NoStop}%
\bibitem [{\citenamefont {McClure}\ \emph {et~al.}(2014)\citenamefont
  {McClure}, \citenamefont {Prins},\ and\ \citenamefont
  {Miller}}]{mcclure2014novel}%
  \BibitemOpen
  \bibfield  {author} {\bibinfo {author} {\bibfnamefont {J.~E.}\ \bibnamefont
  {McClure}}, \bibinfo {author} {\bibfnamefont {J.~F.}\ \bibnamefont {Prins}},
  and\ \bibinfo {author} {\bibfnamefont {C.~T.}\ \bibnamefont {Miller}},\
  }\href@noop {} {\bibfield  {journal} {\bibinfo  {journal} {Computer Physics
  Communications}\ }\textbf {\bibinfo {volume} {185}},\ \bibinfo {pages} {1865}
  (\bibinfo {year} {2014})}\BibitemShut {NoStop}%
\bibitem [{\citenamefont {Armstrong}\ \emph {et~al.}(2016)\citenamefont
  {Armstrong}, \citenamefont {McClure}, \citenamefont {Berrill}, \citenamefont
  {R{\"u}cker}, \citenamefont {Schl{\"u}ter},\ and\ \citenamefont
  {Berg}}]{armstrong2016beyond}%
  \BibitemOpen
  \bibfield  {author} {\bibinfo {author} {\bibfnamefont {R.~T.}\ \bibnamefont
  {Armstrong}}, \bibinfo {author} {\bibfnamefont {J.~E.}\ \bibnamefont
  {McClure}}, \bibinfo {author} {\bibfnamefont {M.~A.}\ \bibnamefont
  {Berrill}}, \bibinfo {author} {\bibfnamefont {M.}~\bibnamefont {R{\"u}cker}},
  \bibinfo {author} {\bibfnamefont {S.}~\bibnamefont {Schl{\"u}ter}}, and\
  \bibinfo {author} {\bibfnamefont {S.}~\bibnamefont {Berg}},\ }\href@noop {}
  {\bibfield  {journal} {\bibinfo  {journal} {Physical Review E}\ }\textbf
  {\bibinfo {volume} {94}},\ \bibinfo {pages} {043113} (\bibinfo {year}
  {2016})}\BibitemShut {NoStop}%
\bibitem [{\citenamefont {Armstrong}\ \emph {et~al.}(2017)\citenamefont
  {Armstrong}, \citenamefont {McClure}, \citenamefont {Berill}, \citenamefont
  {R{\"u}cker}, \citenamefont {Schl{\"u}ter}, \citenamefont {Berg} \emph
  {et~al.}}]{armstrong2017flow}%
  \BibitemOpen
  \bibfield  {author} {\bibinfo {author} {\bibfnamefont {R.~T.}\ \bibnamefont
  {Armstrong}}, \bibinfo {author} {\bibfnamefont {J.}~\bibnamefont {McClure}},
  \bibinfo {author} {\bibfnamefont {M.}~\bibnamefont {Berill}}, \bibinfo
  {author} {\bibfnamefont {M.}~\bibnamefont {R{\"u}cker}}, \bibinfo {author}
  {\bibfnamefont {S.}~\bibnamefont {Schl{\"u}ter}}, \bibinfo {author}
  {\bibfnamefont {S.}~\bibnamefont {Berg}},  \emph {et~al.},\ }\href@noop {}
  {\bibfield  {journal} {\bibinfo  {journal} {Petrophysics}\ }\textbf {\bibinfo
  {volume} {58}},\ \bibinfo {pages} {10} (\bibinfo {year} {2017})}\BibitemShut
  {NoStop}%
\bibitem [{\citenamefont {Liu}\ \emph {et~al.}(2018)\citenamefont {Liu},
  \citenamefont {McClure},\ and\ \citenamefont {Armstrong}}]{liu2018influence}%
  \BibitemOpen
  \bibfield  {author} {\bibinfo {author} {\bibfnamefont {Z.}~\bibnamefont
  {Liu}}, \bibinfo {author} {\bibfnamefont {J.~E.}\ \bibnamefont {McClure}},
  and\ \bibinfo {author} {\bibfnamefont {R.~T.}\ \bibnamefont {Armstrong}},\
  }\href@noop {} {\bibfield  {journal} {\bibinfo  {journal} {Physical Review
  E}\ }\textbf {\bibinfo {volume} {98}},\ \bibinfo {pages} {043102} (\bibinfo
  {year} {2018})}\BibitemShut {NoStop}%
\bibitem [{\citenamefont {McClure}\ \emph {et~al.}(2018)\citenamefont
  {McClure}, \citenamefont {Armstrong}, \citenamefont {Berrill}, \citenamefont
  {Schl{\"u}ter}, \citenamefont {Berg}, \citenamefont {Gray},\ and\
  \citenamefont {Miller}}]{mcclure2018geometric}%
  \BibitemOpen
  \bibfield  {author} {\bibinfo {author} {\bibfnamefont {J.~E.}\ \bibnamefont
  {McClure}}, \bibinfo {author} {\bibfnamefont {R.~T.}\ \bibnamefont
  {Armstrong}}, \bibinfo {author} {\bibfnamefont {M.~A.}\ \bibnamefont
  {Berrill}}, \bibinfo {author} {\bibfnamefont {S.}~\bibnamefont
  {Schl{\"u}ter}}, \bibinfo {author} {\bibfnamefont {S.}~\bibnamefont {Berg}},
  \bibinfo {author} {\bibfnamefont {W.~G.}\ \bibnamefont {Gray}}, and\ \bibinfo
  {author} {\bibfnamefont {C.~T.}\ \bibnamefont {Miller}},\ }\href@noop {}
  {\bibfield  {journal} {\bibinfo  {journal} {Physical Review Fluids}\ }\textbf
  {\bibinfo {volume} {3}},\ \bibinfo {pages} {084306} (\bibinfo {year}
  {2018})}\BibitemShut {NoStop}%
\bibitem [{\citenamefont {Armstrong}\ \emph {et~al.}(2012)\citenamefont
  {Armstrong}, \citenamefont {Porter},\ and\ \citenamefont
  {Wildenschild}}]{armstrong2012linking}%
  \BibitemOpen
  \bibfield  {author} {\bibinfo {author} {\bibfnamefont {R.~T.}\ \bibnamefont
  {Armstrong}}, \bibinfo {author} {\bibfnamefont {M.~L.}\ \bibnamefont
  {Porter}}, and\ \bibinfo {author} {\bibfnamefont {D.}~\bibnamefont
  {Wildenschild}},\ }\href {https://doi.org/10.1016/j.advwatres.2012.05.009}
  {\bibfield  {journal} {\bibinfo  {journal} {Advances in Water Resources}\
  }\textbf {\bibinfo {volume} {46}},\ \bibinfo {pages} {55} (\bibinfo {year}
  {2012})}\BibitemShut {NoStop}%
\bibitem [{\citenamefont {Georgiadis}\ \emph {et~al.}(2013)\citenamefont
  {Georgiadis}, \citenamefont {Berg}, \citenamefont {Makurat}, \citenamefont
  {Maitland},\ and\ \citenamefont {Ott}}]{georgiadis2013pore}%
  \BibitemOpen
  \bibfield  {author} {\bibinfo {author} {\bibfnamefont {A.}~\bibnamefont
  {Georgiadis}}, \bibinfo {author} {\bibfnamefont {S.}~\bibnamefont {Berg}},
  \bibinfo {author} {\bibfnamefont {A.}~\bibnamefont {Makurat}}, \bibinfo
  {author} {\bibfnamefont {G.}~\bibnamefont {Maitland}}, and\ \bibinfo {author}
  {\bibfnamefont {H.}~\bibnamefont {Ott}},\ }\href
  {https://doi.org/10.1103/PhysRevE.88.033002} {\bibfield  {journal} {\bibinfo
  {journal} {Physical Review E}\ }\textbf {\bibinfo {volume} {88}},\ \bibinfo
  {pages} {033002} (\bibinfo {year} {2013})}\BibitemShut {NoStop}%
\bibitem [{\citenamefont {Wilkinson}(1986)}]{wilkinson1986percolation}%
  \BibitemOpen
  \bibfield  {author} {\bibinfo {author} {\bibfnamefont {D.}~\bibnamefont
  {Wilkinson}},\ }\href {https://doi.org/10.1103/PhysRevA.34.1380} {\bibfield
  {journal} {\bibinfo  {journal} {Physical Review A}\ }\textbf {\bibinfo
  {volume} {34}},\ \bibinfo {pages} {1380} (\bibinfo {year}
  {1986})}\BibitemShut {NoStop}%
\bibitem [{\citenamefont {Laplace}(1805)}]{laplace1805traite}%
  \BibitemOpen
  \bibfield  {author} {\bibinfo {author} {\bibfnamefont {P.~d.}\ \bibnamefont
  {Laplace}},\ }\href@noop {} {\bibfield  {journal} {\bibinfo  {journal}
  {Gauthier-Villars, Paris}\ } (\bibinfo {year} {1805})}\BibitemShut {NoStop}%
\bibitem [{\citenamefont {Brown}\ and\ \citenamefont
  {Sethna}(2003)}]{brown2003statistical}%
  \BibitemOpen
  \bibfield  {author} {\bibinfo {author} {\bibfnamefont {K.~S.}\ \bibnamefont
  {Brown}}and\ \bibinfo {author} {\bibfnamefont {J.~P.}\ \bibnamefont
  {Sethna}},\ }\href {https://doi.org/10.1103/PhysRevE.68.021904} {\bibfield
  {journal} {\bibinfo  {journal} {Physical review E}\ }\textbf {\bibinfo
  {volume} {68}},\ \bibinfo {pages} {021904} (\bibinfo {year}
  {2003})}\BibitemShut {NoStop}%
\bibitem [{\citenamefont {AlRatrout}\ \emph {et~al.}(2018)\citenamefont
  {AlRatrout}, \citenamefont {Blunt},\ and\ \citenamefont
  {Bijeljic}}]{alratrout2018spatial}%
  \BibitemOpen
  \bibfield  {author} {\bibinfo {author} {\bibfnamefont {A.}~\bibnamefont
  {AlRatrout}}, \bibinfo {author} {\bibfnamefont {M.~J.}\ \bibnamefont
  {Blunt}}, and\ \bibinfo {author} {\bibfnamefont {B.}~\bibnamefont
  {Bijeljic}},\ }\href {https://doi.org/10.1029/2017WR022124} {\bibfield
  {journal} {\bibinfo  {journal} {Water Resources Research}\ } (\bibinfo {year}
  {2018})}\BibitemShut {NoStop}%
\bibitem [{\citenamefont {Dye}\ \emph {et~al.}(2016)\citenamefont {Dye},
  \citenamefont {McClure}, \citenamefont {Adalsteinsson},\ and\ \citenamefont
  {Miller}}]{dye2016adaptive}%
  \BibitemOpen
  \bibfield  {author} {\bibinfo {author} {\bibfnamefont {A.~L.}\ \bibnamefont
  {Dye}}, \bibinfo {author} {\bibfnamefont {J.~E.}\ \bibnamefont {McClure}},
  \bibinfo {author} {\bibfnamefont {D.}~\bibnamefont {Adalsteinsson}}, and\
  \bibinfo {author} {\bibfnamefont {C.~T.}\ \bibnamefont {Miller}},\
  }\href@noop {} {\bibfield  {journal} {\bibinfo  {journal} {Water Resources
  Research}\ }\textbf {\bibinfo {volume} {52}},\ \bibinfo {pages} {2601}
  (\bibinfo {year} {2016})}\BibitemShut {NoStop}%
\bibitem [{\citenamefont {Schl{\"u}ter}\ \emph {et~al.}(2017)\citenamefont
  {Schl{\"u}ter}, \citenamefont {Berg}, \citenamefont {Li}, \citenamefont
  {Vogel},\ and\ \citenamefont {Wildenschild}}]{schluter2017time}%
  \BibitemOpen
  \bibfield  {author} {\bibinfo {author} {\bibfnamefont {S.}~\bibnamefont
  {Schl{\"u}ter}}, \bibinfo {author} {\bibfnamefont {S.}~\bibnamefont {Berg}},
  \bibinfo {author} {\bibfnamefont {T.}~\bibnamefont {Li}}, \bibinfo {author}
  {\bibfnamefont {H.-J.}\ \bibnamefont {Vogel}}, and\ \bibinfo {author}
  {\bibfnamefont {D.}~\bibnamefont {Wildenschild}},\ }\href@noop {} {\bibfield
  {journal} {\bibinfo  {journal} {Water Resources Research}\ }\textbf {\bibinfo
  {volume} {53}},\ \bibinfo {pages} {4709} (\bibinfo {year}
  {2017})}\BibitemShut {NoStop}%
\end{thebibliography}%


%
%

\end{document}